\documentclass[journal]{IEEEtran}

\ifCLASSINFOpdf
\else
   \usepackage[dvips]{graphicx}
\fi
\usepackage{url}

\usepackage{graphicx}
\usepackage{amsmath, amsthm, amsfonts}
\usepackage{algorithm, algpseudocode}
\usepackage{enumerate}

\newcommand\scale{1}

\newcommand{\nullhyp}{$\text{H}_0$}
\newcommand{\althyp}{$\text{H}_1$}

\begin{document}

\title{ Interbeat Interval Filtering}
\author{\.Ilker Bayram\\ibayram@ieee.org

\thanks{The author is with Whoop, Boston, MA.}}
\maketitle

\begin{abstract}
A number of inhibitory and excitatory factors regulate the beating of the heart. Consequently, the inter-beat intervals (IBIs) are not constant but vary around a mean value, even in the absence of external factors like exercise or stress. Various statistics have been proposed to capture the heart rate variability (HRV) to provide a glimpse into this balance. These statistics usually require accurate estimation of IBIs as a first step. However, estimating IBIs accurately can be challenging in practice, especially for signals recorded in ambulatory conditions. We propose a lightweight state-space filter that models the IBIs as samples of an inverse Gaussian distribution with time-varying parameters. We make the filter robust against outliers by adapting the probabilistic data association filter to the setup. We demonstrate that the resulting filter can accurately identify outliers and the parameters of the tracked distribution can be used to compute a specific HRV statistic (standard deviation of normal-to-normal intervals) without further analysis.
\end{abstract}

\section{Introduction}
Heart beats are triggered by the sinoatrial node, that aims to balance inputs from the autonomic nervous system and other external factors \cite{glass}. Due to the dynamic and varying nature of inputs from these sources (e.g., the sympathetic and parasympathetic nervous systems provide excitatory and  inhibitory inputs respectively), even at rest, and in the absence of extrinsic factors like exercise or stress, the inter-beat intervals exhibit variability. The distribution of IBIs over time provides a window to glimpse into the condition of the cardiovascular system \cite{bro05p691,  cos02p89, moj23p279, yan23p321}.

In practice, non-invasive methods of obtaining IBIs usually rely on detecting individual heartbeats from a modality like electrocardiogram (ECG), photophlethysmography (PPG), phonocardiography (PCG). However, these modalities can be prone to motion or muscle artifacts, or, more generally, the signal quality may be poor. This can lead to failure in peak detection, resulting in erroneous IBIs. 
We propose a light-weight algorithm that can track a time-varying IBI \emph{distribution} that is also robust to errors in IBIs.

\subsection*{Related Work}
We combine ideas from heart beat series modeling and robust state-space filtering. We briefly review each subject.

Simple IBI models are based on the integrate and fire model \cite {glass, bro05p691}. These models consider the hitting times for the running integral of an underlying process $m(t)$. Every time $y(t) = \int_0^{t} m(s)\,ds$, reaches an integer, say, a heart beat is said to occur. If we take $m(t)$ to be a random walk with drift, the IBIs follow an inverse Gaussian distribution (see \cite{bro05p691}, Chp.~4 of \cite{glass}, or Section~\ref{sec:IG}). Building on this model, Barbieri et al. \cite{bar05p424} propose to estimate the parameters of the inverse Gaussian distribution using past IBIs. This yields a more accurate (i.e., lower variance) conditional density for the next beat.

A variation is the `integral pulse frequency model' (IPFM) \cite{bay68p257, rom82p503, bai11p642, mat03p334}, where  $m(t)$ is taken as a  bandlimited signal. By performing spectral analysis of a variant of $m(t)$ (e.g., cancelling the DC component, or normalizing by time-varying heart rate), one can obtain further insight about HRV. 

In \cite{bar06p4}, Barbieri et al., assume that the model parameters of \cite{bar05p424} follow a random walk. They propose an adaptive filter to estimate these parameters. This filter implicitly assumes that the input beats are correct. Taking a further step, Citi et al. \cite{cit12p828} propose a beat correction algorithm, which can be used as a preprocessing step. Given the IBI history, the algorithm in \cite{cit12p828} evaluates the current IBI to decide if it is valid, or falls in one of the anomaly classes. Depending on the class, it fixes the interval to produce either a single IBI or multiple IBIs.

There also exist relevant work targeting not necessarily heart beats but more general neural spiking activity. In \cite{smi03p965, ede04p971}, the authors consider Gaussian processes as latent states that drive a point process through its intensity function (the intensity function setting the probability of firing in an infinitesimal interval  -- see e.g. \cite{snyder, Ross_models} for background). Smith and Brown \cite{smi03p965} propose estimating the parameters of the intensity function as well as the underlying latent space by resorting to the expectation maximization (EM) algorithm  \cite{dem77p1}. In contrast, Eden et al. \cite{ede04p971}  derive recursive update equations that estimate the latent space. In both papers \cite{smi03p965, ede04p971}, the observations are not restricted to spike locations, but are arbitrary points in time. The latent process can, in principle, be estimated for every time instant.

Another category is rule-based algorithms as in \cite{lip19p173, ran07p946}. These algorithms can adapt their behavior to the underlying sequence of inter beat intervals and  aim to fix the inter beat intervals that look anomalous, using simple rules. 
On the other hand, the literature on robust state space filters is vast. However, many papers assume a state and observation space that is (well approximated by) a Gaussian process. Usually, outliers are handled by either incorporating an outlier generation model or manipulating the state or observation distributions to follow a heavy-tailed distribution -- see for instance \cite{wan18p236, rot13p770, aga12p024} for recent work. These approaches are effective, but do not directly address our problem, where the states are clearly non-Gaussian.

An important category is the probabilistic data association filter (PDAF), which is proposed to handle clutter or missing detections \cite{bar09p82, shalom}. PDAF postulates multiple hypotheses to describe the observation generation process. Given observations, one computes the probability of each hypothesis and updates the filter parameters accordingly. In many variants of PDAF, it is also usually assumed that the underlying states follow a Gaussian process.

\subsection*{Contribution}

The technical novelty of the proposed method is the adaptation of the PDAF framework to a filtering problem that involves a non-Gaussian state and observation space. The adopted model implicitly assumes that locally, the IBIs are sampled independently from an inverse Gaussian distribution. This is  arguably less accurate than modified models that take into account correlation across a history, as in \cite{bar05p424}. However, our results indicate that for the purpose of HRV estimation via RMSSD, this is sufficient. This observation is of practical importance, as it allows the proposed filter to efficiently track HRV over time.

Generally, the IBI \emph{distribution} changes much more slowly than individual IBIs. In turn, the parameters describing the distribution can be communicated at a much lower rate. For a battery-powered device, this means considerable compression and savings in terms of either memory, or energy. In addition, from the pseudo-code in Algorithm~\ref{algo:IBI}, we see that the filter involves simple operations. These imply that  the filter is suitable for running on edge devices.  

\begin{algorithm}
\begin{algorithmic}[1]
\Require $\lambda_n, \mu_n | r_{:n} \sim  f_c(\lambda_n, \mu_n | \theta_n)$ per \eqref{eqn:prior}, new IBI $r_{n+1}$, parameters $\gamma$, $p_e$, $\lambda_e$, $\theta_n := \begin{bmatrix}a \, b \, c\,d \end{bmatrix}$
\Ensure Distribution of $\mu_{n+1}, \lambda_{n+1}$ approximated as $f_c(\lambda_{n+1}, \mu_{n+1} | \theta_{n+1})$, parameterized through $\theta_{n+1}$.
\State $\mu^* \gets 2 a /b$
\State $\lambda^* \gets (2ac - b^2) / (2ad)$
\State $h_0 \gets p_e\,\lambda_e\,\exp(-\lambda_e\,r_{n+1})$
\State $h_1 \gets (1-p_e)\,\sqrt{\dfrac{\lambda^*}{2\pi\,r_{n+1}^3}}\,\exp\left(-\dfrac{\lambda(r_{n+1} - \mu)^2}{2 \mu^2\,r_{n+1}}\right)$
\State $\beta_1 \gets h_1 / (h_0 + h_1)$ \Comment{Prob($r_{n+1}$ is not anomalous)}
\State $\beta_0 \gets 1 - \beta_1$ \Comment{Prob($r_{n+1}$ is anomalous) - for reporting}
\State $\theta_{n+1} \gets \gamma \,\theta_n + \beta_1 \begin{bmatrix}0.5 r_{n+1}\\1\\0.5 / r_{n+1}\\ 0.5\end{bmatrix}$
\end{algorithmic}
\caption{IBI Filter\label{algo:IBI}}
\end{algorithm}

\subsection*{Outline}
We start with some background information on the inverse Gaussian distribution in Section~\ref{sec:background}. The filter is derived in Section~\ref{sec:derivation}. Section~\ref{sec:exp} provides numerical experiments that demonstrate the utility of the filter. Section~\ref{sec:conc} contains concluding remarks and outlook.

\section{Background on the Inverse Gaussian Distribution}\label{sec:background}

The assumption underlying the proposed filter is that, locally, IBIs are generated by independently sampling an inverse Gaussian random variable (see \cite{bro05p691} for a motivation, \cite{bar05p424} for a modified model). The filter keeps track of this distribution. 

We provide an overview of the inverse Gaussian distribution (see, e.g., \cite{Ross_models} for derivations of these properties), and describe a useful conjugate prior.

\subsection{The Inverse Gaussian Distribution}\label{sec:IG}

An inverse Gaussian random variable is the time it takes for a random walk with drift to reach unity.
Assuming $w(t)$ denotes a standard Brownian motion process, a random walk with drift can be obtained as
\begin{equation*}
x(t) = \frac{1}{\mu} t + \frac{1}{\sqrt{\lambda}}\,w(t).
\end{equation*}
Here, the term $t / \mu$ ensures that $x(t)$ has linearly increasing mean, whereas $\lambda$ controls its variance. 
The first time that $x(t)$ reaches unity is an inverse Gaussian random variable, parameterized by $\mu$ and $\lambda$. Its pdf is of the form
\begin{equation*}
f_{\text{IG}}(t) = 
\begin{cases}
0, & \text{if } t < 0, \\
\sqrt{\dfrac{\lambda}{2\pi t^3}} \, \exp\left( - \dfrac{\lambda \, (t - \mu)^2}{2 \mu^2\,t} \right), &\text{if }t \geq 0.
\end{cases}
\end{equation*}

The mean and variance of an inverse Gaussian random variable $X$ are
\begin{align}\label{eqn:stat_ig}
\mathbb{E}(X) &= \mu,\\
\text{var}(X) &= \dfrac{\mu^3} { \lambda }.
\end{align}
\subsubsection*{Likelihood for an Inverse Gaussian}
Suppose $x_1, \ldots, x_n$ are independent samples from an inverse Gaussian distribution. The likelihood is of the form
\begin{multline*}
\text{L}(\lambda, \mu; x_1, \ldots, x_n) \\= z\, \lambda^{n/2}\exp\left( - \lambda \biggl( \mu^{-2}\sum_{i=1}^n \frac{x_i}{2}\, - \mu^{-1}\,n + \sum_{i=1}^{n} \frac{1}{2\,x_i}\biggr)\right),
\end{multline*}
where $z$ is independent of $\lambda$ or $\mu$.

\subsubsection*{Natural Conjugate Prior}
The IBI filter operates by updating statistics for $\mu$ and $\lambda$. That is, we retain some ``prior" information about these parameters, and update the prior information as we receive new observations. For Bayesian filtering, this prior information typically consists of a distribution. Adapting to our context, we need to postulate a distribution for $\lambda$, $\mu$, and update that distribution with each new IBI observation. A convenient distribution to use as prior is the natural-conjugate prior \cite{FinkConjugate}.  

From the likelihood function, we find that the natural conjugate is of the form (following the notation of \cite{Palmer1973})
\begin{equation}\label{eqn:prior}
f_c(\lambda, \mu | a, b, c, d) \propto \lambda^{d}\,\exp(-\lambda (a \mu^{-2} -b \mu^{-1} + c)),
\end{equation}
valid for $\mu > 0$, $\lambda > 0$. Palmer \cite{Palmer1973} shows that this function is not integrable in its natural domain $R_+ = \{(\mu, \lambda) : \mu > 0 ,\lambda > 0 \}$. Therefore, it is not a proper distribution. Nevertheless, it becomes integrable if we restrict the region of integration \cite{Palmer1973}. In the following, we will assume this restriction is in place, and employ $f_c(\lambda, \mu)$ to encode the information on $\lambda$, $\mu$. 

Comparing the natural conjugate prior's parameters to the likelihood function, we reach a useful interpretation of the parameters as
\begin{itemize}
\item 2$a$: sum of IBIs
\item $b$ : number of IBIs
\item 2$c$ : sum of the inverse of IBIs
\item 2$d$: number of IBIs 
\end{itemize}

For what follows, we also note that the maximizer of $f_c(\lambda, \mu)$ is (which can be found by setting the gradient to zero),
\begin{align}\label{eqn:mode_fc}
\mu^* &= \frac{2a}{b},\\
\lambda^* &= \frac{c - b^2/(2a)}{d}.
\end{align}

\section{Derivation of the Filter}\label{sec:derivation}

We follow the standard schema for Bayesian filtering \cite{sarkka}.
Specifically, we propose a state-space model where the parameters of the model are allowed to vary with time. Given the set of past IBIs $r_{:n} = \{r_n, r_{n-1}, \ldots\}$, we assume that 
\begin{equation*}
p(\mu_n, \lambda_n | r_{:n}) = f_c(\mu_n, \lambda_n | \theta_n),
\end{equation*}
where $\theta_n = \{a_n, b_n, c_n, d_n\}$ collects together the parameters.

\subsection{State Update Model}
Before receiving the next IBI, we \emph{postulate} that the state updates as
\begin{align*}
p(\mu_{n+1}, \lambda_{n+1} | r_{:n}) &\propto f_c(\mu_{n+1}, \lambda_{n+1} | \theta_n)^{\gamma}\\
&= f_c(\mu_{n+1}, \lambda_{n+1} | \gamma\,\theta_n),
\end{align*}
where $0< \gamma < 1$.

For a unimodal function like $f_c$, raising to a power less than unity preserves the mode, and  increases the variance of the distribution.
This is precisely the behavior that the standard random walk model in Gaussian state-space filtering achieves. For a detailed justification of this specific update model, we refer to \cite{smi79p375}.

Finally, given the distribution $p(\mu_{n+1}, \lambda_{n+1} | r_{:n})$, the next IBI $r_{n+1}$ is assumed to be independent of previous IBIs, and follows the distribution $f_{\text{IG}}(\cdot | \lambda_{n+1}, \mu_{n+1})$.

\subsection{Filter Updates}
Given this model, we can write down the rules for updating the parameters. Specifically, since
\begin{multline*}
p(\mu_{n+1}, \lambda_{n+1} | r_{:n+1}) \\ \propto p(r_{n+1}| \mu_{n+1}, \lambda_{n+1})\cdot p(\mu_{n+1}, \lambda_{n+1} | r_{:n}),
\end{multline*}
we obtain
\begin{align}
p(\mu_{n+1}, &\lambda_{n+1} | r_{:n+1}) \nonumber \\ &\propto f_{\text{IG}}(r_{n+1} | \mu_{n+1}, \lambda_{n+1})\, \cdot f_{c}( \mu_{n+1}, \lambda_{n+1} | \gamma \theta_n) \nonumber \\
&\propto f_{c}( \mu_{n+1}, \lambda_{n+1} | \theta_{n+1})\label{eqn:posterior}
\end{align}
where
\begin{equation*}
\theta_{n+1} = \begin{bmatrix}
a_{n+1}\\
b_{n+1}\\
c_{n+1}\\
d_{n+1}
\end{bmatrix}
= 
\gamma\,\begin{bmatrix}
a_{n}\\
b_{n}\\
c_{n} \\
d_{n}
\end{bmatrix}
 + 
\begin{bmatrix}
 0.5\,r_{n+1}\\
 1\\
 0.5/r_{n+1}\\
0.5
\end{bmatrix}.
\end{equation*}

If, at anytime, the mean and variance of the heart rate distribution is required, they can be estimated using the mode of $f_c(\lambda_n, \mu_n | \theta_n)$, along with \eqref{eqn:stat_ig}
\begin{align*}
\mathbb{E}(\text{HR}) &= \frac{2 a_n}{ b_n}\\
\text{var}(\text{HR}) &= \frac{16\, a_n^4\,d_n}{ b_n^3\, (2ac -b^2)}.
\end{align*}

\subsection{Filtering Out Anomalous Intervals}
The simple tracker outlined above assumes that IBIs do not contain any outliers. In practice, IBI estimation is far from ideal, especially in ambulatory conditions. There can be failures due to incorrect peak detection, which can happen if the input signal quality is poor.
Apart from peak detector failure, it is also possible to observe unexpected IBIs, due to ectopic beats.
We would like the filter to be robust against such IBIs. For the purposes of this document, we do not distinguish ectopic beats from erroneous IBIs, and collect them under the same label, ``anomalous".

We follow the PDAF \cite{bar09p82} framework to derive an algorithm that handles such cases. 

Given each IBI, we have to evaluate two hypothesis:
\begin{itemize}
\item $\text{H}_0^{n+1}$ : $(n+1)^{\text{st}}$ IBI is anomalous
\item $\text{H}_1^{n+1} := \text{H}_0^c$ : $(n+1)^{\text{st}}$ IBI is not anomalous
\end{itemize}
In the following, for simplicity of notation, we will drop the superscript, and write \nullhyp, \althyp.

We are interested in computing the probabilities of these hypotheses. To do that, we declare
\begin{enumerate}[(i)]
\item a prior probability $p_e$ for getting an anomalous IBI,
\item a probability distribution for anomalous observations \label{post:prob}.
\end{enumerate}
For \eqref{post:prob}, we postulate that anomalous observations follow an exponential distribution:
\begin{equation*}
f_e(x) = \lambda_e\, \exp(-\lambda_e\,x)\, u(x),
\end{equation*}
where $u(x)$ denotes the step function.
To incorporate PDAF, we introduced two new scalars, $p_e$, $\lambda_e$, as parameters to tune.

The posterior distribution of $\mu_{n+1}$, $\lambda_{n+1}$ takes the form
\begin{equation}\label{eqn:posterior_mixture}
\sum_{i=0}^1\,p(\mu_{n+1}, \lambda_{n+1} | H_i, r_{:n+1} )\cdot p(H_i| r_{:n+1} )
\end{equation}
For what follows, for $i=0,1$, we define
\begin{align*}
\alpha_i &:= p(\mu_{n+1}, \lambda_{n+1} | H_i, r_{:n+1} ),\\
\beta_i &:=p(H_i| r_{:n+1} ).
\end{align*}

\subsubsection*{Computing Probability of Parameters Given Hypotheses}
$\alpha_i$'s are relatively easy to compute.
Under \nullhyp, we simply take
\begin{equation*}
\alpha_0 = f_{c}( \lambda_{n+1}, \mu_{n+1} | \gamma \theta_n).
\end{equation*}
Under \althyp, we take
\begin{equation*}
\alpha_1 = f_{c}( \lambda_{n+1}, \mu_{n+1} | \theta_{n+1}),
\end{equation*}
whose expression was provided in \eqref{eqn:posterior}.

\subsubsection*{Computing Hypothesis Probabilities} 

We first write
\begin{align*}
\beta_i = p(H_i| r_{:n+1} ) &\propto p(H_i, r_{n+1} | r_{:n} )\\
&= p(r_{n+1} | H_i, r_{:n} )\, p(H_i | r_{:n})
\end{align*}
Whether the $(n+1)^{\text{st}}$ IBI will be anomalous or not does not depend on previous observations.
Therefore,
\begin{equation*}
p(H_i | r_{:n}) = p(H_i) = \begin{cases}
p_e, &\text{for }i=0,\\
1- p_e, &\text{for }i=1,
\end{cases}
\end{equation*}

Consider now the other term. We have 
\begin{equation*}
p(r_{n+1} | H_0, r_{:n} ) = \lambda_e\, \exp( - \lambda_e\,r_{n+1}),
\end{equation*}
and
\begin{equation*}
p(r_{n+1} | H_1, r_{:n} ) = \int\,
f_{\text{IG}}(r_{n+1} | \lambda, \mu)\, \cdot f_{c}( \lambda, \mu | \gamma \theta_n)\, d\lambda\,d\mu.
\end{equation*}
The latter integral does not have a closed form expression. While it's not the only choice, we opt for the simple approximation of 
\begin{equation*}
f_{c}( \lambda, \mu | \gamma \theta_n) = \delta(\lambda - \lambda^*_{n+1})\cdot \delta(\mu- \mu^*_{n+1}),
\end{equation*}
where the maximum likelihood estimates of $\lambda$ and $\mu$ are given in \eqref{eqn:mode_fc}. Plugging those in, we have
\begin{equation*}
p(r_{n+1} | H_1, r_{:n} ) = f_{\text{IG}}(r_{n+1} | \lambda^*_{n+1}, \mu^*_{n+1}).
\end{equation*}
To summarize, we have
\begin{equation*}
\beta_0 = \dfrac{b_0}{b_0 + b_1},\,\,\beta_1 = \dfrac{b_1}{b_0 + b_1},
\end{equation*}
where
\begin{align*}
b_0 &= p_e\,\lambda_e\, \exp( - \lambda_e\,r_{n+1}),\\
b_1 &= (1-p_e)\,f_{\text{IG}}(r_{n+1} | \lambda^*_{n+1}, \mu^*_{n+1}))
\end{align*}

\subsection{Collapsing the Mixture}
We provided expressions to compute the posterior probability in \eqref{eqn:posterior_mixture}. However, this is a mixture distribution. In order to keep the recursions in the same form, we need to reduce the mixture to a single component as is typically done in Gaussian sum filtering and variants \cite{bar09p82}.

To come up with a reduction scheme, we recall the interpretation of the parameters in $\theta_n$. These are either the number of observations, or the sum of a function of the observations. Under \nullhyp, it is as if we miss an observation, whereas in \althyp, we do make the proper observation. Therefore we can simply use a convex combination to combine the two set of parameters. The weight of the convex combination can be taken as $\beta_0$. This leads to the update rule
\begin{align*}
\theta_{n+1} &= \beta_0 (\gamma \theta_n) + \beta_1 \left(\gamma\, \theta_n + v(r_{n+1})\right)\\
&= \gamma\,\theta_n + \beta_1\,v(r_{n+1})
\end{align*}
where
\begin{equation*}
v(r_{n+1}) = \begin{bmatrix}
 0.5\,r_{n+1}\\
 1\\
 0.5/r_{n+1}\\
0.5
\end{bmatrix}.
\end{equation*}
The final expression suggests that we have made a `partial' observation (of weight $\beta_1$) -  an interpretation that is consistent with the motivation for the rule.

Putting it all together gives  Algorithm~\ref{algo:IBI}.

\section{Experiments}\label{sec:exp}
We consider IBI sequences obtained from the Fantasia database \cite{iye96p078} made available through Physionet \cite{gol00p215}. We conduct two experiments to assess different features of the proposed filter, namely anomaly detection, and HRV tracking.

In the first experiment, we compare how the anomaly probability of the proposed filter compares against that of the robust Kalman filter proposed in \cite{wan18p236}.

The second experiment  evaluates the proposed filter's ability to accurately and robustly estimate HRV dynamically. We compare the filter against a two-stage scheme, that comprises a rule-based IBI fixing algorithm followed by a sliding window standard deviation computation. These are compared to ground truth HRV computed also using a sliding window.

\begin{figure}
\centering
\includegraphics[scale = \scale]{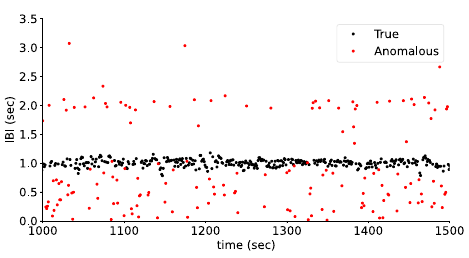}
\caption{True IBIs, and anomalous IBIs obtained by perturbing heart beat locations with error probability $p=0.1$. \label{fig:noisy_obs}}
\end{figure}

For our experiments, given beat locations, we introduce two types of errors:
\begin{itemize}
\item \textbf{Missed detections}, by randomly removing a small percentage, $p_m$, of the beats,
\item \textbf{False detections}, by adding a small percentage, $p_f$, of beats, where each addition is uniformly distributed over the entire collection.
\end{itemize}
Each `missed detection' replaces two normal IBIs with a large anomalous IBI, whereas a `false detection' divides a normal IBI into a consecutive pair of low anomalous IBIs. We set $p_f = p_m=p$ in our experiments.

Fig.~\ref{fig:noisy_obs} shows a small excerpt from `\texttt{f1y02}' (young subject with relatively high HRV) from the Fantasia dataset, with true and anomalous IBIs used in the experiment.

\subsection*{Anomaly Detection Performance}
In order to evaluate the anomaly detection performance, 
for each record in the database,  we add noise to the beat locations with $p = 0.075$\footnote{This is the value where the methods in the next experiment start to break down.}. 
In practice, we will not have access to a probability of anomaly, so we initialized our filter by setting the probability of error to $0.09$. In fact, the value $p$ used in IBI generation and the probability of error in our filter actually have different meanings, so we are not really committing an inverse crime \cite{wir04inverse}. We also ran the IBIs through the robust Kalman filter \cite{wan18p236}. We modified the parameters so as to get as favorable a distribution as possible.

For each noisy beat obtained from the record, we computed the probability that the beat is anomalous using the two methods above. Given the true label (i.e., whether the beat is truly anomalous or not), we can therefore produce two histograms for each method, associated with the estimated anomaly probability when the beat is (i) anomalous, (ii) true. Repeating this over the whole dataset, we obtain the histograms shown in Fig.\ref{fig:histograms}.

\begin{figure}
\centering
\includegraphics[scale = \scale]{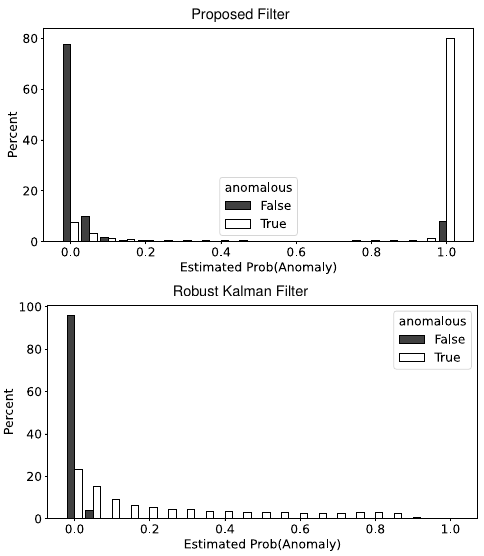}
\caption{Histograms of anomaly probabilities computed via the proposed filter, and the robust Kalman filter from \cite{wan18p236} \label{fig:histograms}.}
\end{figure}

\begin{figure}
\centering
\includegraphics[scale = \scale]{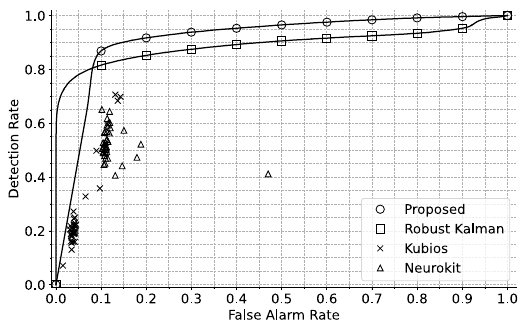}
\caption{ROC curves obtained by thresholding the anomaly probabilities obtained from the proposed filter and the robust Kalman filter \cite{wan18p236} \label{fig:ROC}.}
\end{figure}

Since the two methods produce probabilities, we can use a threshold $\tau \in [0,1]$ to decide whether a beat is true or anomalous. As we vary the threshold, we get different probabilities of false alarm and true detection. The resulting receiver operating characteristic (ROC) curves for the two methods are shown in Fig.~\ref{fig:ROC}.

We also included a comparison using two rule-based methods made available in the \texttt{NeuroKit2} python package \cite{mak21p689}. Because these methods return a binary label, we cannot produce an ROC curve by varying a parameter. Instead, we marked the operating point for each record in the dataset to understand where they stand with respect to the proposed filter and the robust Kalman filter. 

Overall, even though the histograms obtained from the robust Kalman filter do not look nearly as favorable as those of the proposed filter, its ROC curve, especially for low false alarm rates, exhibits almost ideal behavior. However, for a downstream task like HRV estimation, a high anomaly detection rate is arguably more critical because outliers can throw off common statistics like RMSSD or SDNN. In that regard, the proposed filter performs noticeably better, achieving a higher detection rate for slightly higher probability of false alarm (i.e., $\text{False Alarm Rate} > 0.1$).

We think the main reason for the relatively poor performance of the robust Kalman filter is that its assumptions regarding the data distribution is not valid for this scenario. In other words, by tailoring the filter to a more accurate but still simple data model (namely, inverse Gaussian), we are able to retain simplicity of the filter without sacrificing performance.

\subsection*{HRV Accuracy}

To test HRV accuracy, we vary $p$ from 0 to 0.3 in our experiments. $30\%$ error is an exaggeration, and we do not expect methods to handle such error levels successfully. But testing with high error helps us understand how robust the methods under investigation are.

We compare our algorithm against the rule-based algorithm proposed by Lipponen and Tarvainen \cite{lip19p173}, implemented in \texttt{NeuroKit2} \cite{mak21p689}, as well as \texttt{NeuroKit2}'s own method. These algorithms take as input a set of noisy IBIs, determine which IBIs are incorrect, and fix them. 

\begin{figure}
\centering
\includegraphics[scale = \scale]{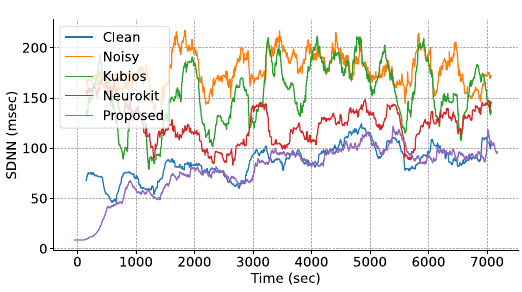}
\caption{\label{fig:HRV trace} Standard deviation of IBIs obtained from the original IBI sequence, the noisy IBI sequence, the tracked distribution from the proposed filter and the fixed sequence of IBIs obtained, from the methods Kubios and Neurokit. The error percentage for this example was set to $p=10\%$. Despite fixing, the std estimates remain considerably high compared to the original.} 
\end{figure}

For the $k^{\text{th}}$ beat at $t_k$, we compute an HRV value by considering the standard deviations of IBIs that fall in a 5-minute window centered around $t_k$ (i.e., the SDNN method). This is done for the original (clean) sequence of IBIs,  the sequence of ``fixed' IBIs, and the noisy IBIs. In addition, we compute the std of the empirical estimate of the inverse Gaussian distribution, namely  $\sqrt{\mu(t)^3 / \lambda(t)}$. The resulting HRV curves are shown for one of the records in the database, \texttt{f1y01} in Fig.~\ref{fig:HRV trace}. We see that the proposed filter accurately tracks the underlying HRV, while the other methods overestimate the HRV even though the probability of outliers is not too high.

\begin{figure}
\centering
\includegraphics[scale = \scale]{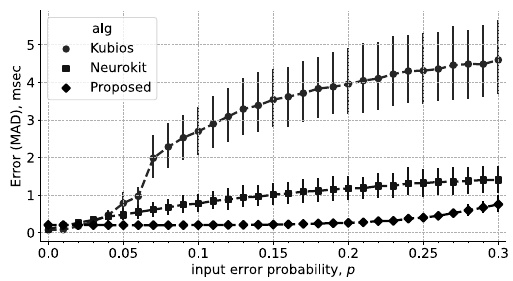}
\caption{\label{fig:MAD} Median absolute deviation between HRV estimates obtained by fixing the noisy beat locations (red) and the proposed filter (black).}
\end{figure}

Given estimates  for each probability of error, $p$,  and record, we compute an accuracy for each method by computing the median absolute deviation (MAD) of the estimated curve and the true HRV curve obtained from the original sequence of IBIs. The resulting distance curves (small being better) with respect to probability of error, $p$, is shown in Fig.~\ref{fig:MAD}

To check how the distribution varies over time, we show the estimated distribution from the proposed filter at different instances in Fig.~\ref{fig:histograms}. For these histograms, we consider 200 consecutive beats. Even though the histogram itself can be noisy due to the low number of beats that contribute to the histogram, the inverse Gaussian distribution appears to correctly track the distribution, leading to a more interpretable summary.

\begin{figure}
\centering
\includegraphics[scale = \scale]{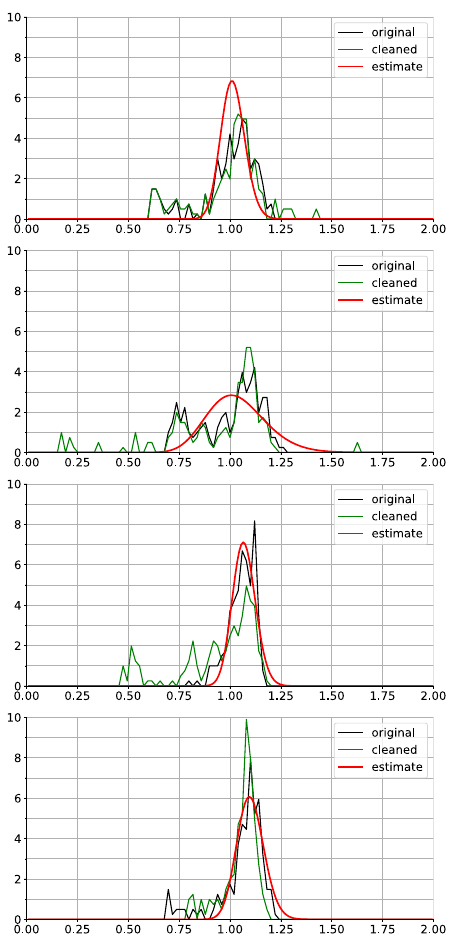}
\caption{\label{fig:histograms} Histograms of IBIs obtained from 200 beats at various instances, compared to the instantaneous estimate of the IBIs from the proposed filter.}
\end{figure}

\section{Conclusion}\label{sec:conc}
This paper introduces a filter that tracks IBIs that is robust to outliers, and  fills a gap in the literature: 
\begin{itemize}
\item it is lean enough to be implemented on a wearable device without requiring too much memory or computation, in contrast to other Bayesian-filtering approaches,
\item it is robust and accurate enough that it performs better than a two-stage approach with a beat-correction stage.
\end{itemize}
The proposed filter builds on a PDAF framework with two hypotheses: the input IBI is an anomaly or not. We plan to investigate a more rigorous version with additional hypotheses for future work.

\end{document}